\documentclass[fleqn,usenatbib,useAMS]{mnras}

\usepackage{graphicx}	% Including figure files
\usepackage{amsmath}	% Advanced maths commands
\usepackage{amssymb}	% Extra maths symbols
\usepackage{multicol}        % Multi-column entries in tables
\usepackage{bm}		% Bold maths symbols, including upright Greek
\usepackage{pdflscape}	% Landscape pages

\usepackage[T1]{fontenc}
\usepackage{ae,aecompl}
\usepackage{orcidlink}

\usepackage{newtxtext,newtxmath}
\title[Polarized QED Cascades over Pulsar Polar Caps]{Polarized QED Cascades over Pulsar Polar Caps}

\author[H.-H. Song and M. Tamburini]{Huai-Hang Song\orcidlink{0000-0002-2587-4658}$^{1}$\thanks{Contact e-mail: \href{mailto:huaihangsong@sjtu.edu.cn}{huaihangsong@sjtu.edu.cn}}
\thanks{Present address: Key Laboratory for Laser Plasmas (MOE), School of Physics and Astronomy, Shanghai Jiao Tong University, Shanghai 200240, China}
and Matteo Tamburini\orcidlink{0000-0002-2897-9826}$^{2}$\thanks{Contact e-mail: \href{Matteo.Tamburini@mpi-hd.mpg.de}{Matteo.Tamburini@mpi-hd.mpg.de}}%
\\
$^{1}$Institute of Physics, Chinese Academy of Sciences, Beijing 100190, China\\
$^{2}$Max-Planck-Institut f{\"u}r Kernphysik, Saupfercheckweg 1, D-69117 Heidelberg, Germany}

\date{Last updated 2024 April 18; in original form 2024 January 18}

\pubyear{2024}

\begin{document}
\label{firstpage}
\pagerange{\pageref{firstpage}--\pageref{lastpage}}
\maketitle

\begin{abstract}
The formation of $e^\pm$ plasmas within pulsar magnetospheres through quantum electrodynamics (QED) cascades in vacuum gaps is widely acknowledged. This paper aims to investigate the effect of photon polarization during the QED cascade occurring over the polar cap of a pulsar. We employ a Monte Carlo-based QED algorithm that accurately accounts for both spin and polarization effects during photon emission and pair production in both single-particle and particle-in-cell (PIC) simulations. Our findings reveal distinctive properties in the photon polarization of curvature radiation (CR) and synchrotron radiation (SR). CR photons exhibit high linear polarization parallel to the plane of the curved magnetic field lines, whereas SR photons, on average, demonstrate weak polarization. As the QED cascade progresses, SR photons gradually dominate over CR photons, thus reducing the average degree of photon polarization. Additionally, our study highlights an intriguing observation: the polarization of CR photons enhances $e^\pm$ pair production by approximately 5\%, in contrast to the inhibition observed in laser-plasma interactions. Our self-consistent QED-PIC simulations in the corotating frame reproduce the essential results obtained from single-particle simulations.
\end{abstract}


%%%%%%%%%%%%%%%%%%%%%%%%%%%%%%%%%%%%%%%%%%%%%%%%%%

%%%%%%%%%%%%%%%%% BODY OF PAPER %%%%%%%%%%%%%%%%%%

% The MNRAS class isn't designed to include a table of contents, but for this document one is useful.
% I therefore have to do some kludging to make it work without masses of blank space.
%\begingroup
%\let\clearpage\relax
%\tableofcontents
%\endgroup
%\newpage

\section{Introduction} \label{introduction}

Pulsars are commonly understood to be rotation-powered neutron stars, and treated as rotating magnetic dipoles characterized by superstrong magnetic fields at the $10^{12}$~G level close to their surface \citep{hewish1968nature, gold1968nature, michel1982rmp, philippov2022araa}. In vacuum, the rapid rotation of pulsars would induce a strong electric field $E_\parallel$ aligned with the magnetic field. The force associated to this induced electric field can surpass the gravitational pull, leading to the extraction of charged particles from the pulsar surface. To resolve this issue, \citet{goldreich1969apj} proposed the presence of a magnetosphere surrounding the pulsar, comprising corotating plasmas that effectively screen the parallel electric field. These plasmas are thought to be $e^{\pm}$ plasmas sustained by quantum electrodynamics (QED) cascades, with the polar cap identified as one of the key regions for their occurrence \citep{sturrock1971apj, ruderman1975apj, arons1979apj}.

\begin{figure}
\includegraphics[width=\columnwidth]{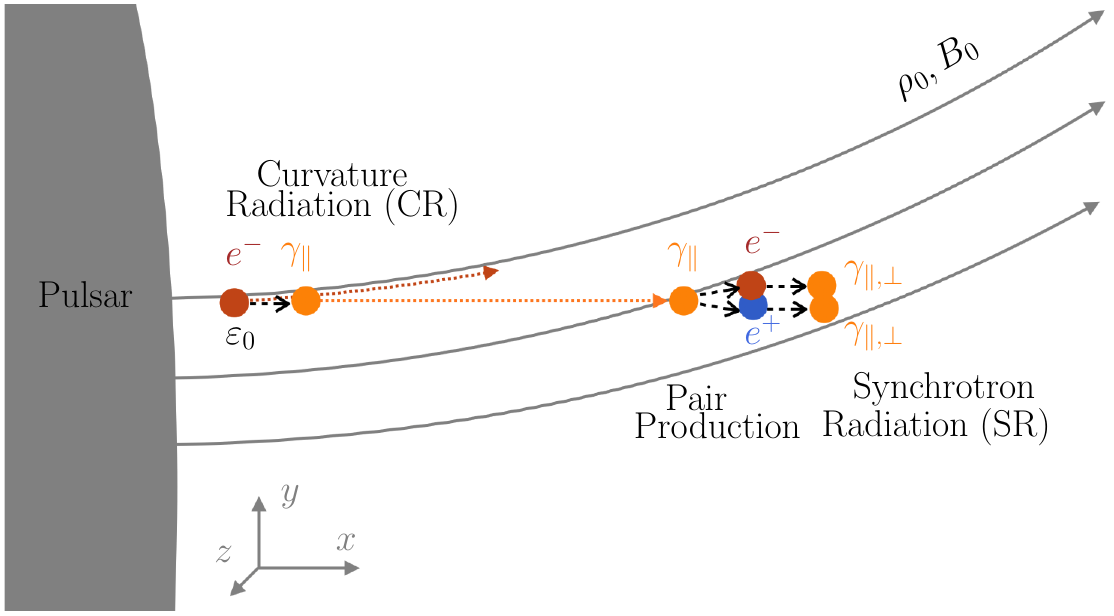}
 \caption{Schematic representation of the polarized QED cascade over a pulsar polar cap. High-energy $e^{\pm}$ particles, drifting in the curved magnetic field, emit $\gamma_\parallel$ photons through CR, which subsequently convert into $e^{\pm}$ pairs. The resulting $e^{\pm}$ pairs can emit both $\gamma_\parallel$ and $\gamma_\perp$ photons via SR. The symbols $\gamma_\parallel$ and $\gamma_\perp$ denote photons polarized parallel and perpendicular to the plane of the magnetic field lines, respectively.} \label{fig1}
\end{figure}

The polarized QED cascade process over a pulsar polar cap is illustrated in Fig.~\ref{fig1}. High-energy $e^{\pm}$, accelerated by the electric field, are confined to move along the curved magnetic field lines with minimal pitch angles, emitting $\gamma$ photons through curvature radiation (CR). These emitted $\gamma$ photons travel straight, leading to an increase in their propagation angles relative to the magnetic field lines until they decay into $e^{\pm}$ pairs. The created $e^{\pm}$ pairs, inheriting large pitch angles from their parent $\gamma$ photons, undergo synchrotron radiation (SR) while simultaneously experiencing strong radiation reaction. Subsequently, SR photons with sufficient energy also decay into $e^{\pm}$ pairs. The unscreened parallel electric field $E_\parallel$ accelerates the created $e^{\pm}$ pairs, reducing their pitch angles and potentially bringing them into the CR regime. This cycle of photon emission and pair production sustains the QED cascade. Ultimately, the generated $e^{\pm}$ plasma becomes sufficiently dense to screen the parallel electric field $E_\parallel$, thus creating the pulsar magnetosphere and approaching a force-free environment.

Magnetohydrodynamics simulations \citep{spitkovsky2006apj} and particle-in-cell (PIC) simulations based on heuristic QED models \citep{philippov2014apjl, chen2014apjl, belyaev2015mnras, cerutti2015mnras, kalapotharakos2018apj, Timokhin2019, CruzPoP2022} have been conducted to replicate force-free pulsar magnetosphere features \citep{contopoulos1999apj}. However, to accurately capture local dynamics like the QED cascade, first-principles QED-PIC simulations that incorporate precise rates of photon emission and pair production are essential. Previous 1D QED-PIC simulations in the corotating frame suggest that the QED cascade over the pulsar polar cap is dynamic \citep{timokhin2010mnras, Timokhin2013}, contrary to previous assumptions of stationarity. This nonstationary creation of $e^{\pm}$ plasmas may hold significance in explaining the origins of pulsar radio emission \citep{philippov2020prl, CruzAJL2021, Bransgrove_2023}. 

Prior QED-PIC simulations investigating QED cascades in a pulsar magnetosphere typically considered unpolarized photon emission and pair production rates, neglecting the effect of $e^{\pm}$ spin and $\gamma$ photon polarization. With recent advancements in strong-field QED physics largely stimulated by the construction of PW-class laser facilities worldwide \citep{danson2019hplse}, there is a growing interest in $e^{\pm}$ spin and $\gamma$ photon polarization effects in various contexts, particularly in laser-electron collisions and laser-plasma interactions \citep{li2019prl, ChenPRL2019, WenPRL2019, wan2020prr, GongPRL2021, song2021njp, song2022prl, GongPRL2023a, GongPRL2023b}. In a pulsar, the directionality of the slightly curved magnetic field above the pulsar polar cap implies the potential significance of spin and polarization effects for the development of QED cascades. In fact, the motion of $e^{\pm}$ is primarily along the magnetic field line, offering a unique perspective compared to studies in storage rings or lasers, where motion is predominantly perpendicular to the external field \citep{jackson1976rmp, wan2020prr, seipt2021njp, song2021njp}. This suggests that QED cascades in pulsars may exhibit distinct spin or polarization properties.

In this paper, we employ a semiclassical Monte Carlo-based single-particle code and a first-principles QED PIC code that resolves $e^{\pm}$ spin and $\gamma$ photon polarization to simulate the polarized QED cascade over a pulsar polar cap. Our simulation results show that emitted CR photons exhibit significant linear polarization parallel to the plane of the curved magnetic field line. This polarization arises from the $e^{\pm}$ acceleration, 
which determines the polarization of the emitted photons, being predominantly directed along the magnetic field plane. In contrast, the fast gyration motion of $e^{\pm}$ with nearly uniformly rotated acceleration directions results in a weak average polarization of SR photons. As CR photons, chiefly emitted by the primary $e^{\pm}$, are more energetic than SR photons, the average photon polarization increases with higher photon energy. The high degree of polarization of CR photons leads to a higher pair-production rate compared to the unpolarized one, resulting in approximately 5\% more $e^{\pm}$ pairs in simulations accounting for photon polarization. Both ``shower-type'' (S-type), i.e. without parallel electric field $E_\parallel$, and self-sustained ``avalanche-type'' (A-type), i.e. with $E_\parallel$, QED cascades exhibit similar features, with A-type cascades having a larger proportion of SR photons due to accelerated secondary $e^{\pm}$ pairs reaching higher energies. Fully 3D single-particle simulation results are confirmed by 1D QED-PIC simulations in the corotating frame.

The remaining sections of this paper are organized as follows. Section \ref{method} provides an overview of the strong-field QED theory and simulation method utilized. In Section \ref{analysis}, we conduct a detailed analysis of photon polarization in three distinct radiation regimes using single-particle simulations. Section \ref{cascade} delves into the comprehensive cascade of photon emission and pair production, employing both single-particle simulations and QED-PIC simulations. Specifically, we study the time evolution of photon emission and analyze the impact of photon polarization on pair production. Finally, Section \ref{summary} offers a concise summary of our key findings.

\begin{figure*}
\includegraphics[width=1\textwidth]{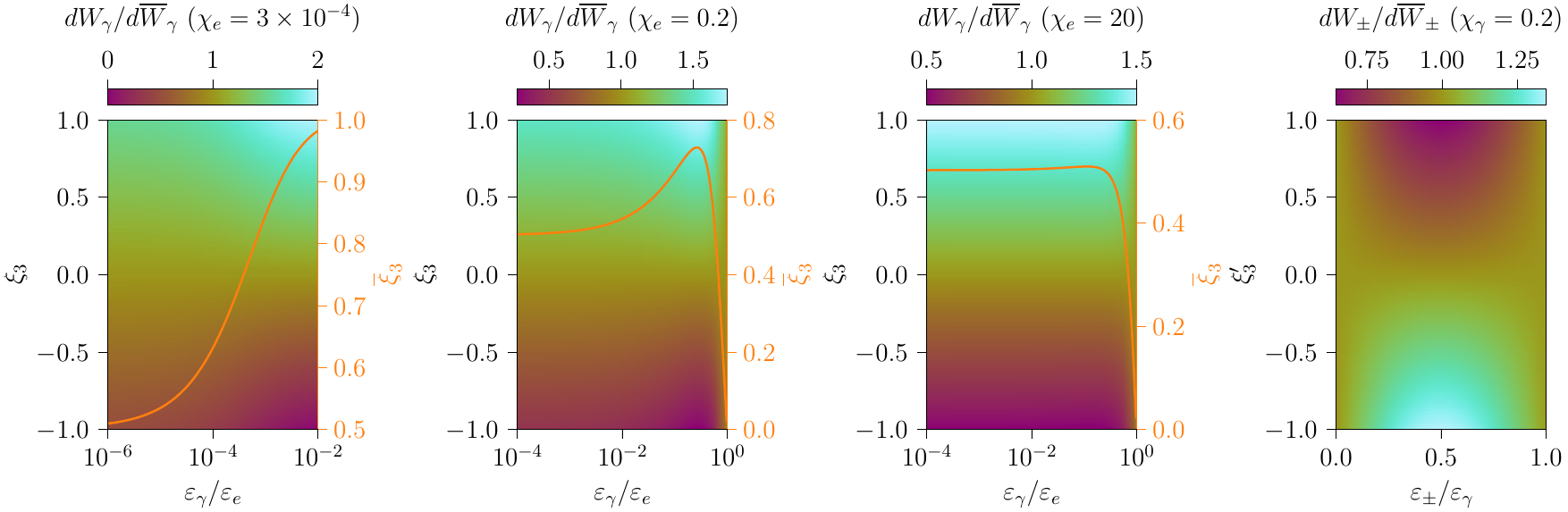}
 \caption{The normalized rates of photon emission, $W_\gamma/\overline{W}_\gamma$ (three left-most panels), and pair production, $W_\pm/\overline{W}_\pm$ (right-most panel), are calculated using Eqs.\eqref{eq1} and \eqref{eq2}, where $\overline{W}_\gamma$ and $\overline{W}_\pm$ are the corresponding unpolarized rates. In the three left panels, the orange line illustrates the average photon polarization, $\overline{\xi}_3$, as a function of the photon energy ratio, $\varepsilon_\gamma/\varepsilon_e$. The QED parameters in these panels are set as $\chi_e=3\times10^{-4}$, $\chi_e=0.2$, and $\chi_e=20$, while the right-most panel displays results with $\chi_\gamma=0.2$. In the right-most panel, observe the symmetry with respect to the $\varepsilon_\pm/\varepsilon_\gamma=0.5$ axis, and the increased probability of particle creation with $\xi'_3<0$.} \label{fig2}
\end{figure*}

\section{Simulation method} \label{method}

Our implementation of polarized QED modules remains consistent across the single-particle and the QED-PIC codes. We utilize semiclassical spin- and polarization-resolved rates for photon emission and pair production in the locally-constant-field approximation (LCFA) \citep{ritus1985jslr,baier1998qed}. To preserve fully 3D polarization information, we choose the mean probability axis as the quantization axis \citep{Cain}. A detailed algorithm description and benchmarking can be found in the work of one of the authors \citep{song2022prl}. In this section, we summarize the aspects relevant for this study.

In a pulsar polar cap the velocity of primary particles responsible for CR emission is nearly aligned with the magnetic field lines. Consequently, in accordance with the strong-field QED theory of polarization-dependent pair production (as outlined below), the spin of the created $e^{\pm}$ pairs is basically \emph{perpendicular} to the direction of the magnetic field\footnote{This stands in stark contrast to scenarios such as storage rings or laser-plasma interactions, where the velocity of primary particles is typically transverse to the external field, and the spin of the created $e^{\pm}$ pairs is predominantly \emph{aligned} with the external field.}. Hence, the fast spin precession of $e^{\pm}$ within the superstrong magnetic field of the pulsar results into a rapid depolarization. Indeed, in our fully spin- and polarized-resolved simulations, we observe no significant net $e^{\pm}$ spin polarization. Consequently, our focus shifts to the analysis of photon polarization. With averaged (initial state) and summed (final states) $e^\pm$ spin but resolved photon polarization, the semiclassical rates for photon emission and pair production in the LCFA can be written as \citep{baier1998qed}
\begin{equation}\label{eq1}
\begin{aligned}
\frac{d^2W_{\gamma}}{d\varepsilon_\gamma dt}=&\frac{C_0}{\varepsilon_e^2}\left[\frac{\varepsilon_e^2+\varepsilon_e'^2}{\varepsilon_e\varepsilon_e'}K_{\frac{2}{3}}(y_1)-{\rm Int}K_{\frac{1}{3}}(y_1) + \xi_3K_{\frac{2}{3}}(y_1) \right],
\end{aligned}
\end{equation}
\begin{equation}\label{eq2}
\begin{aligned}
\frac{d^2W_{\pm}}{d\varepsilon_\pm dt}&=\frac{C_0}{\varepsilon_\gamma^2}\left[\frac{\varepsilon_{+}^2+\varepsilon_{-}^2}{\varepsilon_{+}\varepsilon_{-}}K_{\frac{2}{3}}(y_2)+{\rm Int}K_{\frac{1}{3}}(y_2)-\xi_3'K_{\frac{2}{3}}(y_2)\right],
\end{aligned}
\end{equation}
where $C_0=\alpha m_e^2c^4 / (\sqrt{3}\pi \hbar)$, $K_{\nu}(y)$ is the modified Bessel function of the second kind with a non-integer $\nu$, ${\rm Int}K_{\frac{1}{3}}(y) \equiv \int_{y}^{\infty}K_{\frac{1}{3}}(x)dx$, $m_e$ is the electron mass, $c$ is the speed of light in vacuum, $\hbar$ is the reduced Planck constant, and $\alpha\approx 1/137$ is the fine structure constant. In Eq.~\eqref{eq1}, $\varepsilon_e$ and $\varepsilon_e'$ are the $e^{\pm}$ energies before and after the photon emission, respectively, $\varepsilon_\gamma$ the emitted photon energy, $\varepsilon_e'=\varepsilon_e-\varepsilon_\gamma$, and $y_1=2\varepsilon_\gamma/(3\chi_e\varepsilon_e')$. In Eq.~\eqref{eq2}, $\varepsilon_\pm$ and $\varepsilon_\gamma$ are the energies of the created $e^{\pm}$ pair and the decayed $\gamma$ photon, respectively, and $y_2=2\varepsilon_\gamma^2/(3\chi_\gamma\varepsilon_{+}\varepsilon_{-})$. Two strong-field QED parameters $\chi_e=(e\hbar/m_e^3c^4)|F_{\mu\nu}p^{\nu}|$ and $\chi_\gamma=(e\hbar^2/m_e^3c^4)|F_{\mu\nu}k^{\nu}|$ determine the importance of photon emission and pair production, with $\hbar k^{\nu}$ and $p^{\nu}$ being the photon and $e^\pm$ four-momentum, respectively, and $F_{\mu\nu}$ being the electromagnetic field tensor. In our simulations, we set the momentum direction of generated particles and their parent particle to be collinear. 

The Stokes vector, also referred to as photon polarization hereafter, $\bm\xi=(\xi_1,\xi_2,\xi_3)$ of the emitted photon in Eq.~\eqref{eq1}  is defined in the basis ($\bm e_1, \bm e_2, \bm e_v$), where $\bm e_v$ is the unit vector along the photon velocity, $\bm e_1$ is the unit vector along the transverse acceleration of its parent $e^\pm$, and $\bm e_2=\bm e_v \times \bm e_1$. The positive (negative) value of the photon polarization $\xi_3>0$ ($\xi_3<0$) represents that photons are linearly polarized predominantly along the $\bm e_1$ ($\bm e_2$) direction. Another similar Stokes vector $\bm\xi'=(\xi_1',\xi_2',\xi_3')$ in Eq.~\eqref{eq2} is defined in another basis ($\bm e_1', \bm e_2', \bm e_v$), where $\bm e_1'$ is the unit vector along the transverse acceleration direction of the created positron, and $\bm e_2'=\bm e_v \times \bm e_1'$. Here, the photon Stokes vector describes a mixed state, with $|\bm\xi|\leq1$, $|\bm\xi'|\leq1$. In Eqs.~\eqref{eq1} and \eqref{eq2}, we have omitted the $\xi_1$ and $\xi_2$ components because they are zero on average in our case.

The two Stokes components $\xi_3$ and $\xi_3'$ in Eqs.~\eqref{eq1} and \eqref{eq2} can be transformed into each other via \citep{mcmaster1961rmp}
\begin{equation}
\xi_3'=\xi_3\cos(2\phi),
\end{equation}
where $\phi$ is the angle between the two transverse acceleration directions of $\bm e_1$ and $\bm e_1'$. If $\bm e_1$ and $\bm e_1'$ are collinear, then $\phi=0$ or $\pi$, and $\xi_3'=\xi_3$; otherwise, $\xi_3'=-\xi_3$ if $\bm e_1$ is perpendicular to $\bm e_1'$. We will see that the latter case holds for the decay of CR photons.

The preferred polarization of emitted photons can be determined from Eq.~\eqref{eq1}. The polarized photon emission rates $W_\gamma$ normalized to the unpolarized ones $\overline{W}_\gamma$ are displayed in the three left panels of Fig.~\ref{fig2}. Three QED parameters $\chi_e=3\times10^{-4}$, $\chi_e=0.2$, and $\chi_e=20$ are shown, which correspond to CR, SR, and super-SR (defined in the next section), respectively. For all three regimes, the normalized photon emission rate $W_\gamma/\overline{W}_\gamma$ increases with the photon polarization $\xi_3$ for fixed photon energy. This indicates that the emitted photons with $\xi_3>0$ dominate, being mainly linearly polarized along the $\bm e_1$ axis, i.e., the transverse acceleration axis. Note that since $\chi_e$ of CR is much smaller than 1, high-energy photons are still much less energetic than their parent $e^{\pm}$. In each panel, the average photon polarization $\overline\xi_3$ as a function of the photon energy ratio $\varepsilon_\gamma/\varepsilon_e$ is presented by the orange line. For CR, $\overline\xi_3$ grows from 0.5 at $\varepsilon_\gamma/\varepsilon_e\lesssim10^{-6}$ to nearly 1 at $\varepsilon_\gamma/\varepsilon_e\approx10^{-2}$. Thus, the highest energy CR photons are highly linearly polarized. For SR, where the emitted photons can take away most of the energy of their parent $e^{\pm}$, $\overline\xi_3$ also increases from roughly 0.5, reaching a maximum of about 0.7 at $\varepsilon_\gamma/\varepsilon_e\approx0.3$ in the case of $\chi_e=0.2$, and then drops to 0 when $\varepsilon_\gamma/\varepsilon_e$ approaches 1. This is because the photon polarization is highly sensitive to the spin polarization of their parent $e^{\pm}$ when the emitted photon energy is close to the energy of the parent $e^{\pm}$. Therefore, the opposite result to CR photons is concluded: the highest energy SR photons are weakly polarized. For super-SR with a large $\chi_e$, e.g., $\chi_e=20$, the photon polarization peak at middle energies nearly disappears, leading to $\overline\xi_3$ almost monotonically decreasing from 0.5 to 0 with the increase of $\varepsilon_\gamma/\varepsilon_e$. Note that, as it will be clear below, super-SR often occurs for large values of $\chi_e$, but does \emph{not} coincide with $\chi_e\gg1$.

Similarly, the pair production rate also depends on the polarization of parent photons according to Eq.~\eqref{eq2}. The right-most panel of Fig.~\ref{fig2} shows that the normalized pair production rate $W_{\pm}/\overline W_{\pm}$ decreases with the increase of $\xi_3'$ for a fixed $e^{\pm}$ energy, indicating that photons with $\xi_3'<0$ are more prone to decay into $e^{\pm}$ pairs. The effect of photon polarization on the pair production rate can be up to a maximum of about 30\%, peaking at $\varepsilon_\gamma/\varepsilon_e=0.5$ for $\chi_\gamma\lesssim1$.

Note that the photon polarizations $\xi_3$ and $\xi_3'$ in Eqs.~\eqref{eq1} and \eqref{eq2} are defined with respect to different axes, i.e.,  $\bm e_1$ and $\bm e_1'$, respectively. For convenience, hereafter, we use the same basis $({\bm e}_y, {\bm e}_z, {\bm e}_x)$ for $\gamma$ photons propagating along the assumed magnetic field direction ${\bm e}_x$. In the following, with this choice the basis, the positive (negative) value of $\xi_3$ represents linear photon polarization mainly along the $\bm e_y$ ($\bm e_z$) direction.

\section{Photon polarization} \label{analysis}

\begin{figure}
\includegraphics[width=\columnwidth]{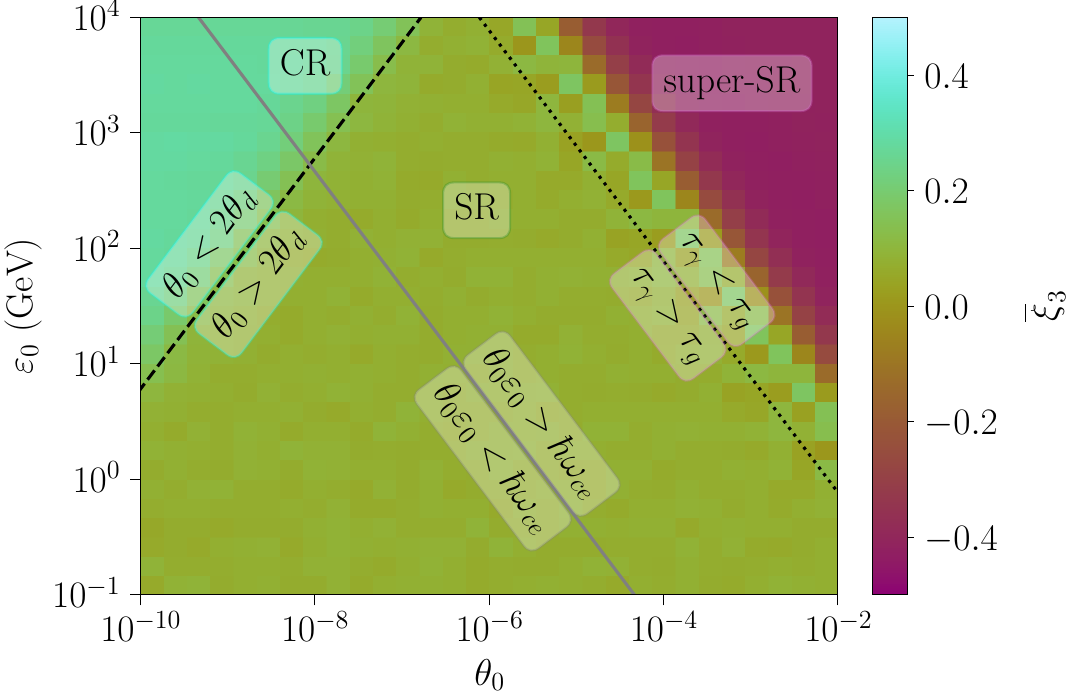}
 \caption{Simulation results reporting the average photon polarization $\overline{\xi}_3$ over a simulation duration equivalent to one photon emission length. The plot illustrates the dependence on the initial pitch angle, $\theta_0$, and energy, $\varepsilon_0$, of the injected electron beam. Three distinct regimes -- CR, SR, and super-SR -- are demarcated by the dashed line $\theta_0=2\theta_d$ and the dotted line $\tau_{\gamma}=\tau_g$. Additionally, the solid line $\theta_0\varepsilon_0=\hbar\omega_{\rm{ce}}$ is presented, signaling the region where Landau energy quantization becomes essential.} \label{fig3}
\end{figure}

\begin{figure*}
\includegraphics[width=1\textwidth]{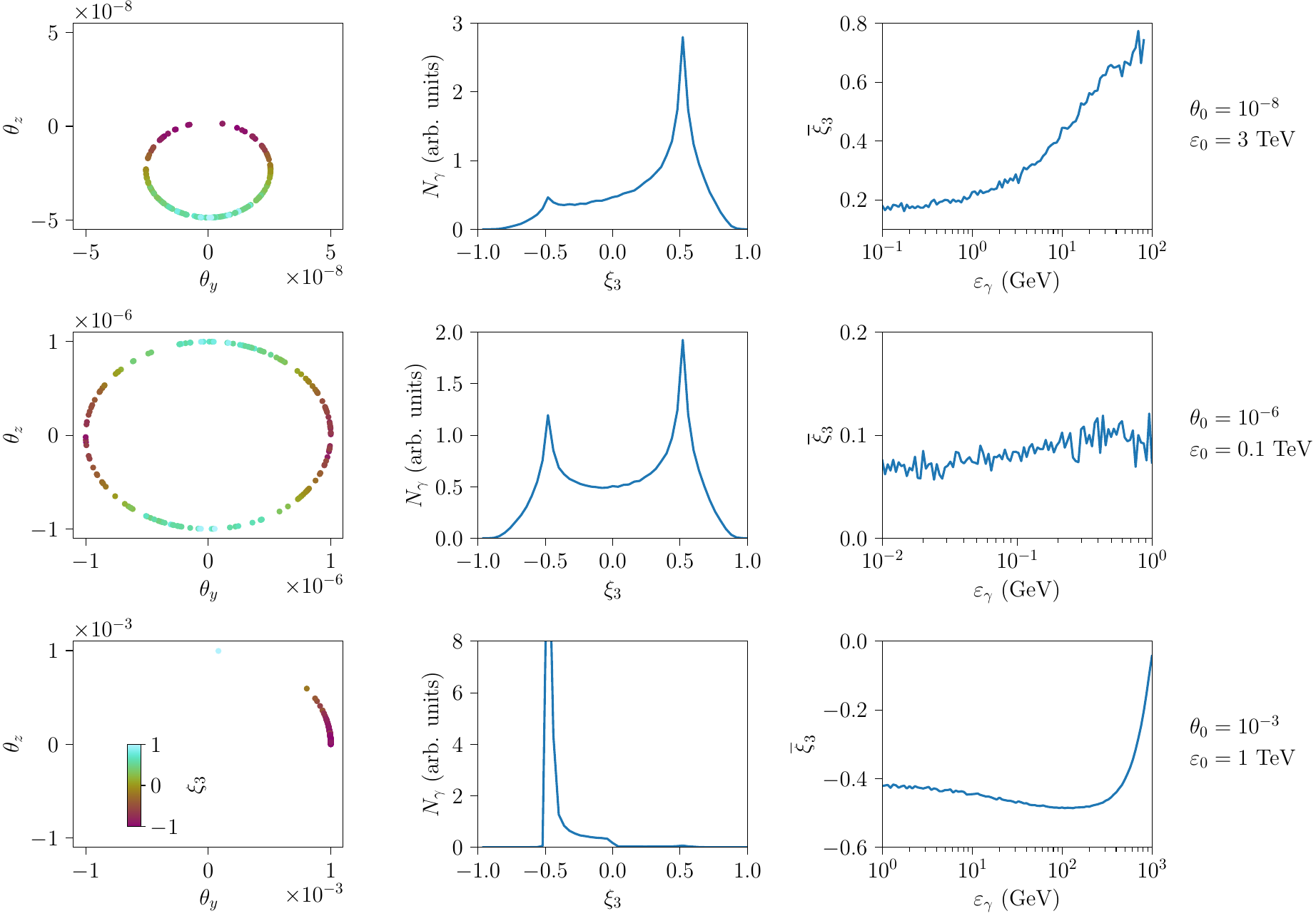}
 \caption{Comparison of photon polarizations in the CR $[\theta_0,\varepsilon_0]=[10^{-8}, 3~\rm TeV]$, SR $[\theta_0,\varepsilon_0]=[10^{-6}, 0.1~\rm TeV]$, and super-SR $[\theta_0,\varepsilon_0]=[10^{-3}, 1~\rm TeV]$ regimes. Left panels: the emitted photons versus their pitch angle components $\theta_y$ and $\theta_z$ at the emission time. The color represents the corresponding photon polarization $\xi_3$. Middle panels: the histograms of the photon polarization $\xi_3$. Right panels: the average photon polarization $\overline{\xi}_3$ as a function of the photon energy $\varepsilon_\gamma$.} \label{fig4}
\end{figure*}

In this section, we focus on a clear comparison of the polarization between CR and SR photons. Specifically, we only consider photon emission within approximately one photon emission length, and pair production is neglected in simulations.

The polarization characteristics of CR and SR, often referred to as synchro-curvature radiation \citep{cheng1996apj, vigano2015mnras}, can be analyzed in a unified manner by considering an electron beam with an initial energy $\varepsilon_0$ and pitch angle $\theta_0$ injected into a purely curved magnetic field. In this case, we assume a simple field configuration where the magnetic field is predominantly directed along the $+\bm e_x$ direction and slightly curved towards the $+\bm e_y$ direction, with a constant magnitude of $B_0=4\times10^{11}$ G. To reduce computational costs, we adopt a constant curvature radius of $\rho_0=10$ km, which is smaller than the typical values (100--1000 km) but comparable to that of offset polar caps \citep{harding2011apj}. We set the initial pitch angle of the electron beam towards the $+\bm e_y$ direction, specifically $\theta_y|_{t=0}=\theta_0$ and $\theta_z|_{t=0}=0$, where $\theta_{y,z}=\arctan(p_{y,z}/|p_x|)$, since photon emission occurs in the plane of the curved magnetic field lines ($x$-$y$ plane), preferentially. 

To confine the charged $e^{\pm}$ moving along the curved magnetic field, the net Lorentz force must act as a centripetal force, pointing from the $e^{\pm}$ position to the curvature center of the magnetic field line. The Lorentz force acting on $e^{\pm}$ is determined by its pitch angle. Therefore, the pitch angle $\theta$ of $e^{\pm}$ must be asymmetric about the magnetic field line during its propagation, a phenomenon known as curvature drift velocity \citep{alfven1963book}. The center of the pitch angle in the $\theta_y$-$\theta_z$ space will drift by $\theta_d=({\rm max}|\theta_z| - {\rm max}|\theta_y|)$, which gives the  the \emph{drift angle} \citep{alfven1963book}
\begin{equation} \label{CR_angle}
\theta_d=\frac{\gamma_0m_ec}{eB_0\rho_0}\approx 3.3\times10^{-9}\frac{\varepsilon_0(\rm TeV)}{B_0(10^{12}~\rm G)\rho_0(10~\rm km)}.
\end{equation}
Photon emission attributed to CR occurs only for $e^\pm$ at small pitch angles $\theta\lesssim2\theta_{\rm d}$, while SR dominates if $\theta \gg2\theta_{\rm d}$. The defined boundary between these two mechanisms is essentially the same as that based on radiation power \citep{cheng1996apj, vigano2015mnras}, differing only by an unimportant factor of 2. This is because the radiation power is directly related to the QED parameter $\chi_e\approx\gamma_0\theta B_0/B_{\rm cr}$, which is proportional to $\theta$ at $\theta\ll1$, where $B_{\rm cr}=m_e^2 c^2/e \hbar \approx 4.41\times10^{13}$~G is the QED critical strength of the magnetic field. 

Finally,  when considering photon polarization, a unique radiation regime, which we call \emph{super-SR}, should be treated separately. This regime is characterized by a photon emission time $\tau_\gamma=1/\int_0^{\varepsilon_e}d\varepsilon_\gamma(d^2W_\gamma/d\varepsilon_\gamma dt)$ smaller than the $e^{\pm}$ gyration period $\tau_{\rm g}=2\pi\gamma_0 m_e/eB_0$, corresponding to high $e^{\pm}$ energy and large pitch angle. To determine the boundary between the SR and super-SR regime, the photon emission time can be approximated as $\tau_\gamma\approx\hbar\gamma_0/(1.44\alpha m_ec^2\chi_e)$ in the weak QED regime of $\chi_e\lesssim0.1$ \citep{baier1998qed}. The consistency of this approximation can be checked a posteriori (see below). Utilizing the small-angle approximation for $\chi_e$, we obtain $\tau_\gamma\approx\hbar B_{\rm cr}/(1.44\alpha m_e c^2\theta B_0)$, which is independent of the particle's energy. The condition $\tau_\gamma<\tau_g$ yields $\gamma_0 \theta > 1/(2.88\pi \alpha)\approx 15$, independent of the magnetic field strength. Using $\tau_\gamma=\tau_g$, the critical QED parameter at the boundary between SR and super-SR is $\chi_e^{\rm cr}= B_0/(2.88\pi\alpha B_{\rm cr})\approx 0.34\times B_0(10^{12}~\rm G)$, independent of the specific pitch angle and electron energy. For $B_0\lesssim10^{12}~\rm G$, this gives $\chi_e^{\rm cr}\lesssim0.1$, consistent with the initial assumption of the boundary between SR and super-SR being in the weak QED regime. Energetic super-SR photons are typically characterized by high linear polarization, but with the polarization direction \emph{perpendicular} to that of CR photons. 

In summary, the three different radiation regimes are
\begin{equation}
\begin{cases}
{\rm CR}: \theta<2\theta_d,\\ 
{\rm SR}: \theta>2\theta_d~{\rm and}~\tau_\gamma>\tau_g,\\
\operatorname{super-SR}: \tau_\gamma<\tau_g.
\end{cases}
\end{equation}
To investigate photon polarization in the three radiation regimes, we conducted a series of 3D single-particle simulations, comprising $30 \times 30$ sets. The simulations involved logarithmic scans of the initial pitch angle $\theta_0$, ranging from $10^{-10}$ to $10^{-2}$, and electron energy $\varepsilon_0$, ranging from $100$~MeV to $10$~TeV. The numerical time step, $\Delta t$, was set to $10^{-4}\tau_\gamma$. Utilizing the Boris scheme for $e^{\pm}$ dynamics, we maintained accuracy in photon emission and CR photon polarization even with a relatively large time step, $\Delta t > \tau_g$ \citep{boris1970conf, qin2013pop}. In fact, test simulations performed by injecting a $\varepsilon_0=1$~TeV electron beam into the curved magnetic field of $B_0=4\times10^{11}$~G and $\rho_0=10$~km, and employing two time-steps $\Delta t=10^{-6}$~ns and $10^{-3}$~ns showed no accumulation of numerical effects and small relative differences. The simulation outcomes, illustrating the average photon polarization $\overline\xi_3$ as a function of $\theta_0$ and $\varepsilon_0$, are presented in Fig.~\ref{fig3}. The sign of $\overline\xi_3$ effectively distinguishes three radiation regimes: $\overline\xi_3 > 0$ for CR, $\overline\xi_3 \approx 0$ for SR, and $\overline\xi_3 < 0$ for super-SR. It is worth noting that while Landau quantization is typically disregarded for magnetic field strengths $B_0/B_{\rm cr} \lesssim 0.1$ \citep{harding1987apj}, the region characterized by very small pitch angles and low energies, defined by $\theta_0\varepsilon_0 < \hbar\omega_{ce}$, is forbidden due to Landau energy level quantization. Here, $\omega_{ce} = eB_0/m_e$ represents the nonrelativistic gyration frequency of an electron in the magnetic field. In the following, our discussion focuses on the parameter space where $\theta_0\varepsilon_0 \gg \hbar\omega_{ce}$. For better understanding, Fig.~\ref{fig4} illustrates three selected cases, each representing a distinct radiation regime.

(i) \emph{CR}. With an initial pitch angle $\theta_0=10^{-8}$ and energy $\varepsilon_0=3$ TeV ($2\theta_d\approx5.0\times10^{-8}$), the photon emission resides in the CR regime. Due to the magnetic field's curvature, the pitch angle is not symmetric about the field line; instead, its center shifts along the $-\bm e_z$ direction by $\theta_{\rm d}$, as shown in the top-left panel of Fig.~\ref{fig4}. Since the QED parameter $\chi_e$, which determines the photon emission rate, is proportional to the pitch angle, most photons are emitted at $\theta_z<0$. According to the polarized photon emission rate, photons are predominantly linearly polarized along the net transverse acceleration direction of $e^{\pm}$. Consequently, on average CR photons are linearly polarized along the $\bm e_y$ direction, resulting in an average photon polarization $\overline\xi_3>0$ in the chosen basis $({\bm e}_y, {\bm e}_z, {\bm e}_x)$. Consistent with the theoretical expectation, $\overline\xi_3$ monotonically increases with the photon energy.

(ii) \emph{SR}. When the initial pitch angle is $\theta_0=10^{-6}$ and the energy is $\varepsilon_0=0.1$ TeV ($2\theta_d\approx1.7\times10^{-9}$), and simultaneously the gyration period $\tau_g$ is smaller than the photon emission time $\tau_\gamma$, the photon emission transitions into the SR regime. A notable difference from CR is that the pitch angle of SR is essentially symmetric about the magnetic field line in the $\theta_y$-$\theta_z$ space, as shown in the middle-left panel of Fig.~\ref{fig4}. Consequently, photons are nearly uniformly emitted during the $e^\pm$ gyration motion. Although photons remain linearly polarized along the acceleration direction, the overall polarization at different pitch angles nearly cancels out. The curvature effect of the magnetic field induces a weak inhomogeneity of the photon emission in the $\bm e_z$ direction, leading to a slight dominance of $\xi_3>0$ photons. The average polarization $\overline\xi_3$ for all photons is approximately 8\%, considerably smaller than the CR polarization degree and almost insensitive to the photon energy. Thus, for computational convenience, we treat SR photons as weakly polarized and directly set the polarization of all SR photons to a constant value of 8\% in our subsequent simulations, which allows us to use a numerical time-step larger than the gyration period.

(iii) \emph{Super-SR}. For $e^{\pm}$ with both large pitch angles and high energies ($\theta_0=10^{-3}$ and $\varepsilon_0=1$ TeV, $2\theta_d\approx1.7\times10^{-8}$), they are likely to emit photons on timescales shorter than the gyration period, simultaneously losing most of their energies due to radiation reaction. In this super-SR regime, photon emission concentrates near the initial $e^{\pm}$ position on the $+\bm e_y$ axis, as shown in the bottom-left panel of Fig.~\ref{fig4}. Consequently, super-SR photons exhibit a distinct linear polarization directed along $\bm e_z$, i.e., the initial transverse acceleration, resulting in $\overline\xi_3<0$ in the $({\bm e}_y, {\bm e}_z, {\bm e}_x)$ basis. This is markedly different from the CR and SR regimes. The polarization degree of super-SR photons decreases to zero at high energy, in agreement with the theoretical expectation for $\chi_e>1$. However, as we will see below, the existence of this super-SR regime is unlikely for $B_0\gtrsim4\times 10^{11}$~G. This is because a relatively large pitch angle corresponds to newly created $e^{\pm}$ pairs, which typically do not possess sufficient energy to emit in the super-SR regime. Consequently, this unique radiation regime will not be further detailed in this work.

\section{Polarized QED cascade} \label{cascade}

\begin{figure*}
\includegraphics[width=\textwidth]{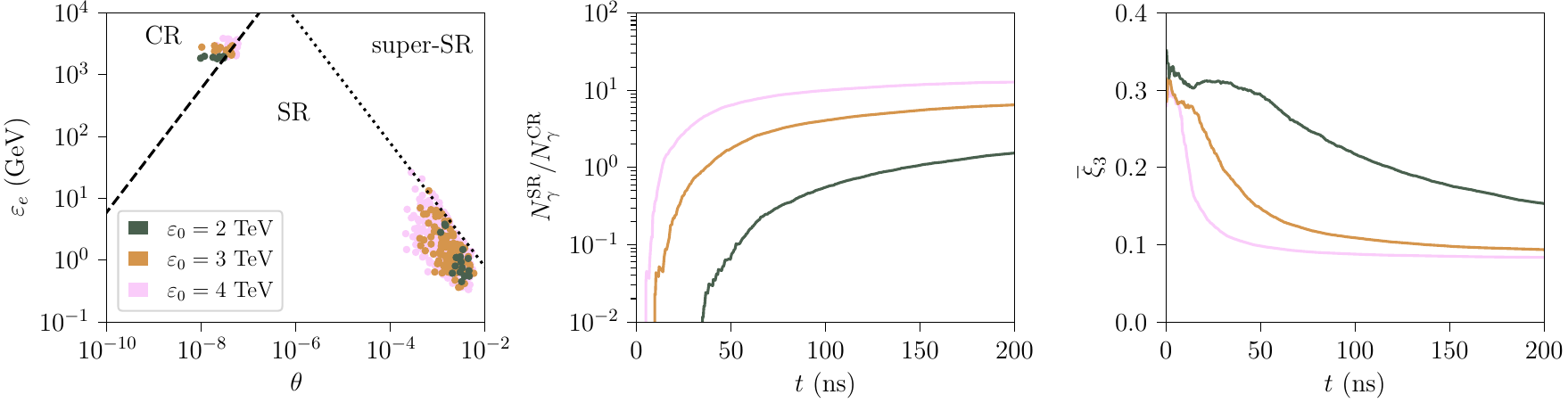}
 \caption{Single-particle simulation results of the S-type cascade for three values of the initial electron beam energy $\varepsilon_0=2$ TeV, 3 TeV, and 4 TeV. Left panel: the emitted photons in the $\theta$-$\varepsilon_e$ space. Middle panel: temporal evolution of the ratio $N_\gamma^{\rm SR}/N_\gamma^{\rm CR}$ between the number of SR photons and CR photons. Right panel: temporal evolution of the average photon polarization $\overline{\xi}_3$.} \label{fig5}
\end{figure*}

\begin{figure*}
\includegraphics[width=\textwidth]{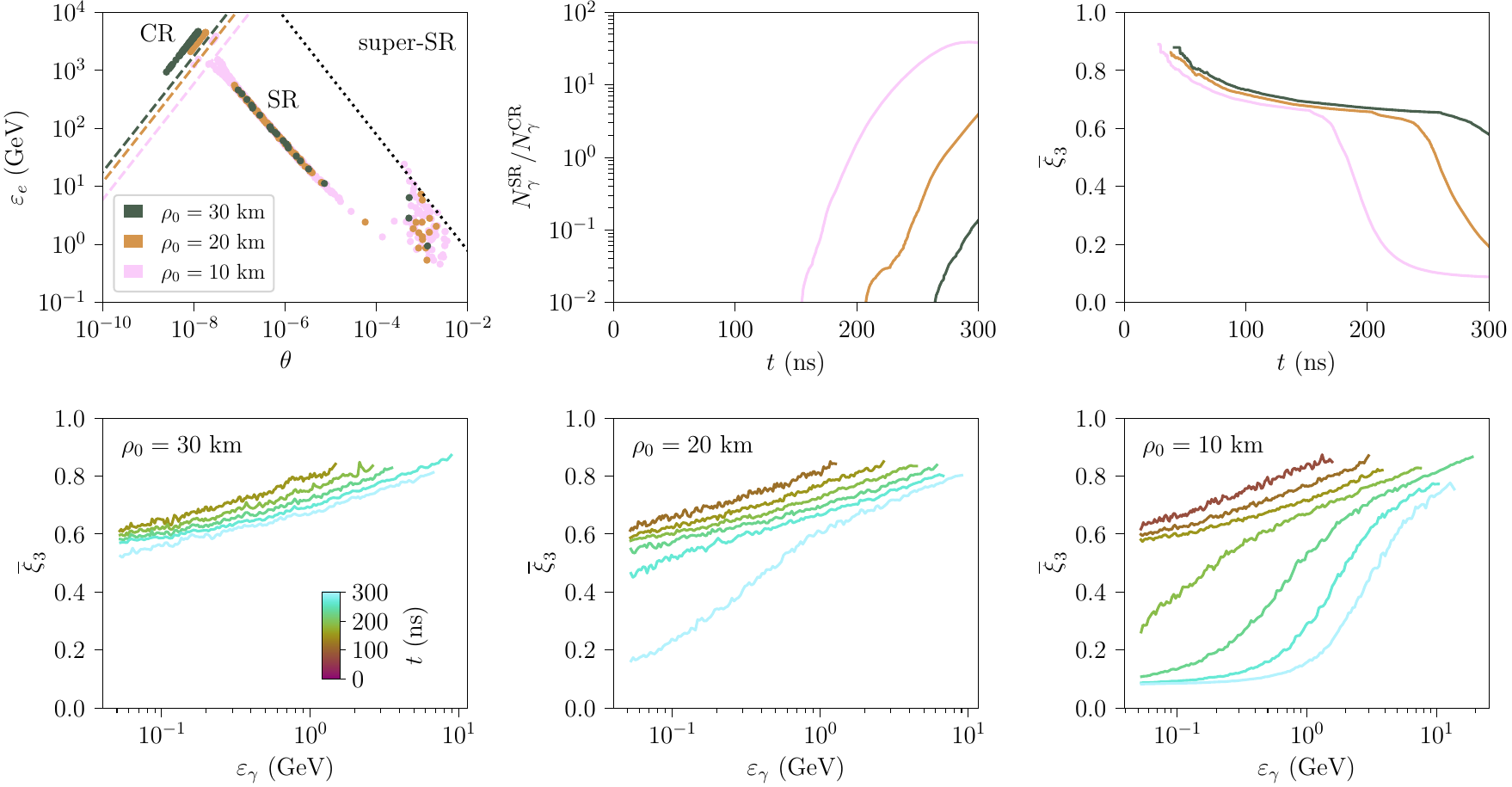}
 \caption{Top panels: the same plots as in Fig.~\ref{fig5} but for the A-type cascade with electrons initially at rest and for three choices of the magnetic field curvature radius, specifically $\rho_0=10$~km, 20~km, and 30~km. Dashed lines, matching the respective colors, delineate boundary lines ($\theta=2\theta_d$) indicating CR and SR regimes. Bottom panels: the temporal evolution of the average photon polarization $\overline{\xi}_3$ versus the photon energy $\varepsilon_\gamma$ across three distinct cases.} \label{fig6}
\end{figure*}

In this section, we investigate the complete process of the polarized QED cascade by considering both photon emission and pair production. Our study encompasses single-particle simulations involving S-type and A-type cascades, with static external fields, as well as first-principles PIC simulations where the parallel electric field is self-consistently incorporated.

\subsection{S-type QED cascades} \label{S-type}

We first study the S-type QED cascade initiated by an energetic electron beam in a pure magnetic field through single-particle simulations. We consider a monoenergetic electron beam with various initial energies, specifically $\varepsilon_0=2$~TeV, 3~TeV, and 4~TeV, tangentially injected into the curved magnetic field. The magnetic field parameters align with those in section \ref{analysis}, with $B_0=4\times10^{11}$~G and $\rho_0=10$~km. In our simulations, the numerical time-step $\Delta t$ is set to be $3.3\times10^{-4}$~ns, which proves sufficient to resolve the dynamics for our cases.

From the simulation results reported in Fig.~\ref{fig5}, we can effectively distinguish the emitted CR and SR photons in the $\theta$-$\varepsilon_e$ space for all three initial electron energies $\varepsilon_0$. As a reminder, $\theta$ and $\varepsilon_e$ represent the pitch angle and energy of the emitting $e^\pm$, respectively. CR photons emerge from injected electrons of high energies ($\varepsilon_e\sim$ TeV) and small pitch angles ($\theta\sim10^{-8}$--$10^{-7}$). On the other hand, SR photons originate from created secondary $e^{\pm}$ pairs with relatively lower energies ($\varepsilon_e\sim$ GeV) and larger pitch angles ($\theta\sim10^{-4}$--$10^{-2}$). The maximum pitch angle of CR equals twice the drift angle $\theta_d$ as defined in Eq.~\eqref{CR_angle}. As $\varepsilon_0$ increases, the pitch angle of SR decreases, because higher-energy CR photons can decay into $e^{\pm}$ pairs with smaller pitch angles. SR photons lie below the boundary line $\tau_\gamma=\tau_g$ of the super-SR regime, resulting in globally weak polarization.

As the QED cascade develops, the proportion of SR photons grows, eventually surpassing CR photons. This trend is evident in the temporal evolution of the number ratio of SR photons to CR photons, $N_\gamma^{\rm SR}/N_\gamma^{\rm CR}$ (see the middle panel of Fig.~\ref{fig5}). The growth rate of $N_\gamma^{\rm SR}/N_\gamma^{\rm CR}$ exhibits significant variability with the initial electron energy $\varepsilon_0$, increasing much more rapidly with a larger $\varepsilon_0$. At the final simulation time $t=200~\rm ns$, $N_\gamma^{\rm SR}\approx N_\gamma^{\rm CR}$ for $\varepsilon_0=2$ TeV, while $N_\gamma^{\rm SR}\gg N_\gamma^{\rm CR}$ for $\varepsilon_0=4$ TeV. The weakly polarized SR photons emitted later contribute to the decrease in the average photon polarization $\overline\xi_3$, depending also on the initial electron energy $\varepsilon_0$ (see the right panel of Fig.~\ref{fig5}).  For $\varepsilon_0=2$ TeV, the effects of pair production are relatively weak, leading to a decrease in average photon polarization from 30\% to 15\%. For larger initial energies, $\overline\xi_3$ rapidly decreases until it converges to a weak SR photon polarization of about 8\%.

\subsection{A-type QED cascade} \label{A_type}

To comprehensively explore the polarized QED cascade, we extend our investigation to include the A-type cascade by incorporating the $e^{\pm}$ acceleration process. In this scenario, an electron beam initially at rest experiences a strong parallel electric field $E_\parallel$ along the magnetic field direction. We set the constant electric field as $E_\parallel=5\times10^{10}$~V/m, consistent with the values obtained from PIC simulations (see below). The magnetic field strength remains fixed at $B_0=4\times10^{11}$~G, and we consider three curvature radii, $\rho_0=10$ km, 20 km, and 30 km.

In the top-left panel of Fig.~\ref{fig6}, we report the emitted photons in the $\theta$-$\varepsilon_e$ space under three curvature radii at the time $t=300$~ns. Unlike the S-type cascade, the low-energy $e^{\pm}$ pairs created here are accelerated to higher energies by the parallel electric field. Consequently, in addition to CR photons from initial electrons ($\theta<10^{-8}$) and SR photons from created $e^{\pm}$ pairs ($\theta>10^{-4}$), we observe additional SR photons emitted by the accelerated $e^{\pm}$ pairs with intermediate pitch angles ($10^{-8}<\theta<10^{-4}$). As in the S-type cascade, super-SR photons are notably absent in all cases.

For the largest curvature radius $\rho_0=30$ km, the QED cascade evolves slowly, with pair production and consequently SR emission occurring toward the end of the simulation. The reduction in photon polarization over time is evident across the entire photon energy range, as depicted in the bottom-left panel of Fig.~\ref{fig6}. For the smallest curvature radius $\rho_0=10$~km, at $t=300$~ns SR photons have already outnumbered CR photons. The average photon polarization $\overline\xi_3$ decreases over time until it stabilizes at about 8\%. The net polarization of low-energy photons experiences a rapid decline, while high-energy photons maintain a noticeably higher level of polarization, as illustrated in the bottom-right panel of Fig.~\ref{fig6}. This is attributed to CR photons, which are highly polarized, having higher energies than SR photons, which are weakly polarized. Therefore, low-energy photons mainly undergo depolarization because of SR emissions. For the intermediate curvature radius, $\rho_0=20$~km, the QED cascade is still rapidly developing at $t=300$~ns, exhibiting polarization characteristics between the cases of $\rho_0=10$~km and $\rho_0=30$~km.

\subsection{Photon polarization effect on pair production} \label{polarized_pair_production}

\begin{figure}
\includegraphics[width=\columnwidth]{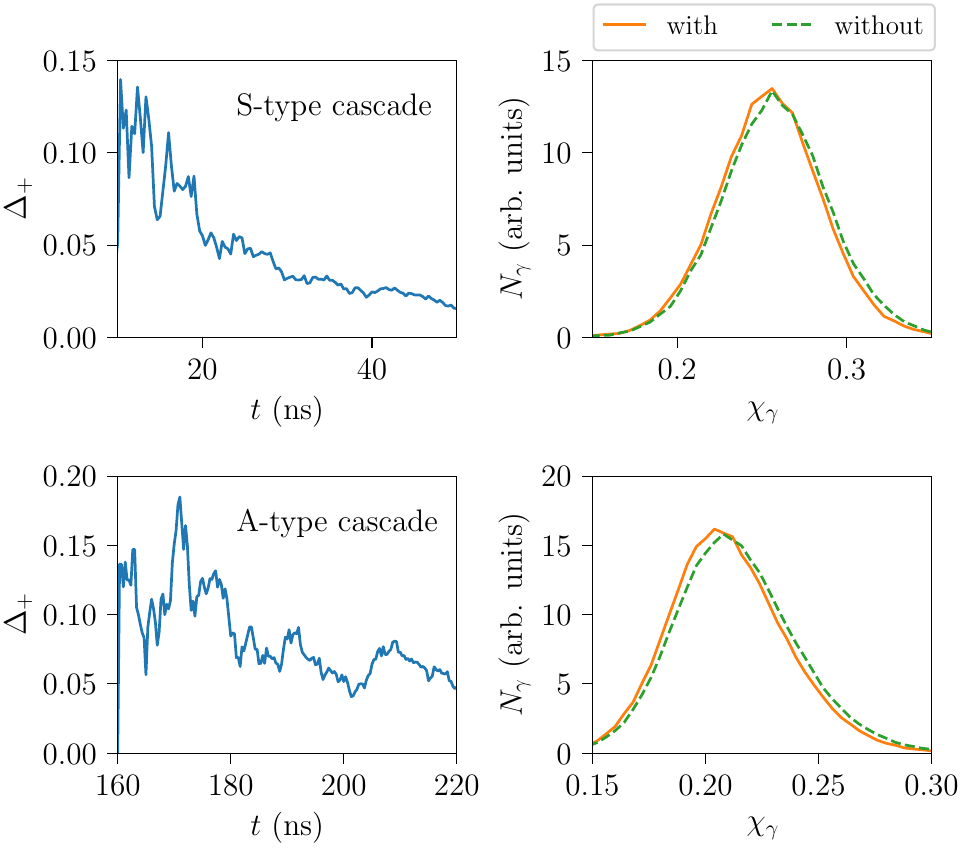}
\caption{Left panels: the temporal evolution of the relative difference of the positron number $\Delta_+=(N_+-\overline{N}_+)/\overline{N}_+$, where $N_+$ and $\overline N_+$ are the number of generated positrons with and without considering the photon polarization, respectively. Right panels: the QED parameter $\chi_\gamma$ of decayed photons with and without the photon polarization. Top panel: S-type cascade with $\varepsilon_0=3$ TeV, and the other parameters as in Fig.~\ref{fig5}. Bottom panel: A-type cascade with $\rho_0=10$~km, and the other parameters as in Fig.~\ref{fig6}.} \label{fig7}
\end{figure}

In this subsection, we investigate the influence of photon polarization on pair production. According to Eq.~\eqref{eq2}, the photon polarization can affect the pair production rate by as much as 30\%. When applying Eq.~\eqref{eq2}, one should pay particular attention to the selected basis for the photon Stokes vector $\bm \xi$. The two basis vectors $\bm e_1$ in Eq.~\eqref{eq1} and $\bm e_1'$ of pair production in Eq.~\eqref{eq2} are mutually perpendicular for CR. This arises from the fact that the net acceleration direction of $e^{\pm}$ in CR is mainly parallel to the plane of magnetic field lines, while for newly created $e^{\pm}$ pairs, the acceleration direction is mainly perpendicular to it. Consequently, this results in $\xi_3\approx-\xi_3'$. From Eq.~\eqref{eq2}, the specific polarization of CR photons leads to a higher pair production rate compared to unpolarized photons. This finding starkly contrasts with results obtained in laser fields, where the photon polarization effect was observed to suppress pair production \citep{wan2020prr, seipt2021njp, song2021njp}. In fact, in laser fields, highly polarized photons are emitted via an SR-like mechanism, leading to $\xi_3\approx\xi_3'$.

The relative difference in the positron number $\Delta_+=(N_+-\overline N_+)/\overline N_+$ between polarized and unpolarized results is reported in Fig.~\ref{fig7}, where $N_+$ and $\overline N_+$ are the number of generated positrons with and without considering the photon polarization, respectively. Figure~\ref{fig7} reveals that the number of created positrons is enhanced by more than 10\% at the initial stage of the cascade due to the high polarization of CR photons. In the S-type cascade, at later times the relative difference narrows to less than 5\%, even though the polarization of CR photons remains high. The comparison of the $\chi_\gamma$ of decayed photons with versus without photon polarization (see the right column of Fig.~\ref{fig7}) reveals that for the unpolarized case, $\chi_\gamma$ at photon decay is slightly larger than in the polarized case. In fact, unpolarized photons with the same energy as polarized photons posses smaller decay rates and travel longer distances, thus acquiring larger pitch angles, and consequently slightly larger $\chi_\gamma$ than polarized photons before decaying into pairs. In the A-type cascade, photon polarization has a similar effect on pair production, as illustrated in the bottom panel of Fig.~\ref{fig7}. The relative difference $\Delta_+$ can exceed 10\% at the initial stage and is also reduced to about 5\% with the development of the QED cascade. Similar to the S-type cascade, a slight difference in $\chi_\gamma$ is observed between polarized and unpolarized results. At a later stage, a key factor to the reduction in the pair yield difference between polarized and unpolarized photons is the domination of unpolarized SR photons over polarized CR photons.

\begin{figure}
\includegraphics[width=\columnwidth]{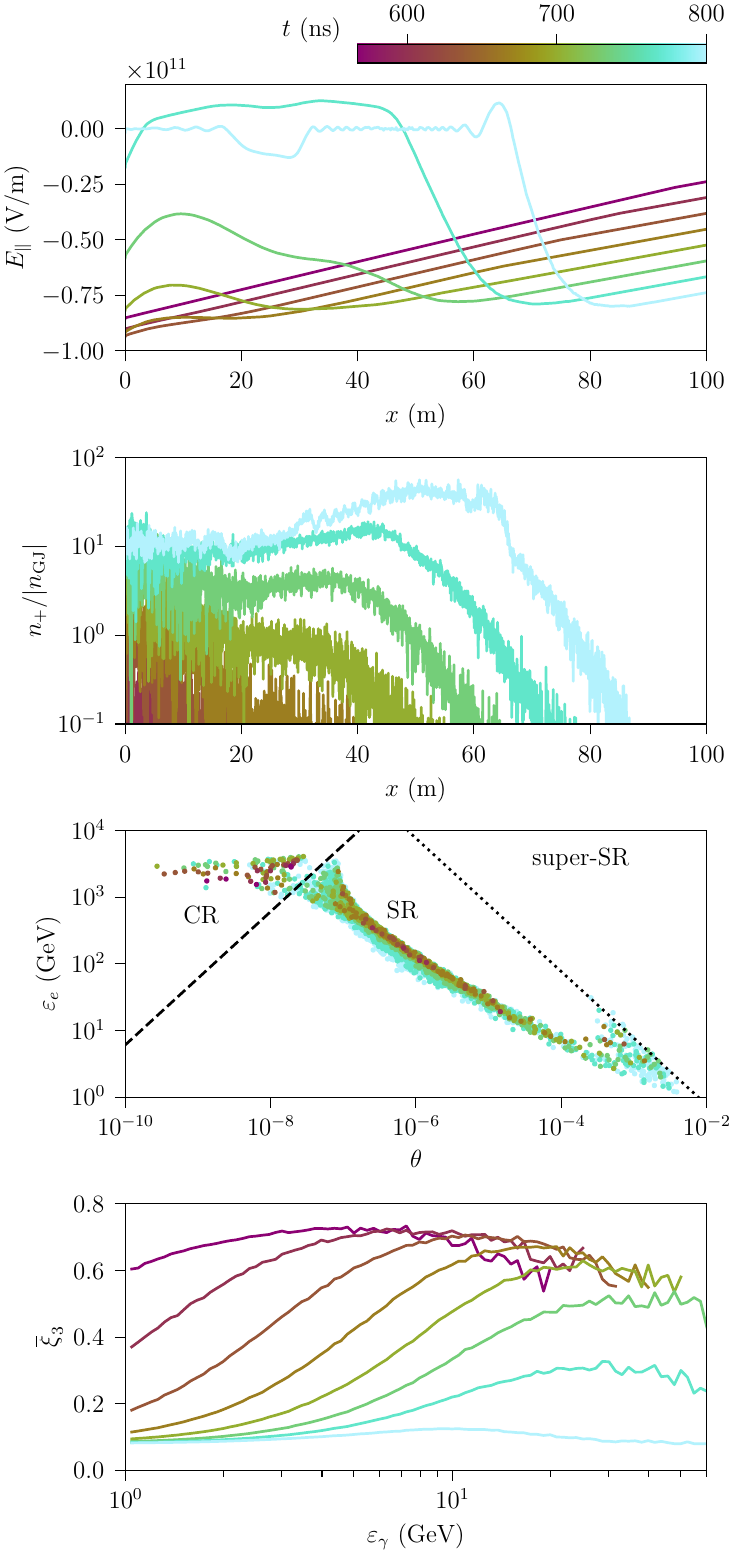}
 \caption{From top to bottom panel: Spatial distributions of the parallel electric field $E_\parallel$ and the normalized positron density $n_+/|n_{\rm GJ}|$. Emitted $\gamma$ photons in the $\theta$-$\varepsilon_e$ space. Temporal evolution of the photon polarization $\overline{\xi}_3$ in relation to the photon energy $\varepsilon_\gamma$.} \label{fig8}
\end{figure}

\subsection{Polarized QED-PIC simulations} \label{polarized_QED_sim}

In our final set of simulations, we employ first-principles 1D (1D geometry but fully 3D particle momentum) QED-PIC simulations in the corotating frame using the {\scshape YUNIC} code \citep{song2021arxiv}. {\scshape YUNIC} is a relativistic PIC code that includes polarized QED modules and has been previously utilized for studying polarized QED cascades in laser-plasma interactions \citep{song2021njp, song2022prl}. Notably, the electric field in these simulations is inhomogeneous and self-consistently solved. This is a crucial distinction from the single-particle simulations, where the electric field is not evolving. Here the electric field is excited by the outflow of charged $e^{\pm}$ plasmas and subsequently screened by the created dense $e^{\pm}$ plasmas, all of which can be captured by the QED-PIC simulation. Our simulation setup is similar to the pioneering work \citep{timokhin2010mnras}. There are several improvements in our QED-PIC code, in which a standard Monte-Carlo QED method developed in recent years by the laser-plasma community is adopted \citep{elkina2011prab, gonoskov2015pre, ridgers2014jcp, tamburini2017, montefiori2023}, and enriched with the polarization effects introduced in section \ref{method}. The magnetic field has a strength of $B_0=4\times10^{11}$~G and a curvature radius of $\rho_0=10$ km. 

In the corotating frame, the governing Goldreich-Julian (GJ) equations are expressed as \citep{goldreich1969apj, timokhin2010mnras}
\begin{equation}
\label{eq6}
\begin{split}
&\nabla\times \bm E=4\pi(n-{n_{\rm GJ}}),\
&\frac{\partial \bm E}{\partial t}=-4\pi(\bm j-{\bm j_{\rm m}}).
\end{split}
\end{equation}
Initially, to satisfy Eq.~\eqref{eq6}, we set the net charge density $n|_{t=0}$ to be the GJ density $n_{\rm GJ} = -\bm\Omega_0\cdot\bm B/2\pi c$ and the net current density $\bm j|_{t=0}$ to be $\bm j_{\rm m}$. Assuming the angular velocity $\bm\Omega_0$ of the pulsar rotation (rotation period $P_0=1$~s) is parallel to its magnetic moment, i.e., $n_{\rm GJ}<0$, the vacuum gap is closed with $\bm E|_{t=0}=0$ at the beginning of the simulation. As the $e^{\pm}$ particles flow away from the simulation zone, Eq.~\eqref{eq6} implies that a strong parallel electric field $E_\parallel$ is excited, accelerating $e^{\pm}$ and triggering the QED cascade.

Simulating the entire QED cascade from the vacuum gap closing to its opening and closing again is challenging due to the exponential growth in particle number. While particle merging methods can alleviate computational pressure, computational particle division only in the position-momentum space might result into loss of polarization information. Thus, we decided to avoid using particle merging, and limited ourselves to simulate the early stage of the QED cascade. Nonetheless, the created $e^{\pm}$ plasma at the end of the simulation is already dense enough to screen the excited electric field. Thus, the obtained PIC simulation results are representative of a well-developed QED cascade. Specifically, we set the initial charge densities of electrons and positrons to be $n_-|_{t=0}=2n_{\rm GJ}$ and $n_+|_{t=0}=-n_{\rm GJ}$, respectively, and their initial current densities to be $\bm j_-|_{t=0}=\bm j_{\rm m}=cn_{\rm GJ}$  and $\bm j_+|_{t=0}=0$. The total simulation box is $L_x=120$~m, with each cell resolved by $d_x=1/384$~m and initially filled with 10 electrons and 10 positrons. Absorbing boundaries are applied at both left and right sides, and no $e^{\pm}$ is supplied from the boundaries, which is similar to the ``anti-GJ'' space-charge limited flow \citep{beloborodov2008apj, timokhin2010mnras}.

With the propagation of the initial flow of $e^\pm$ outside the computational box, a parallel electric field is excited, linearly increasing towards the pulsar surface (left boundary) to accelerate $e^\pm$ to high energies (see the top panel of Fig.~\ref{fig8}). When the QED cascade starts, the density of created $e^{\pm}$ plasma near the pulsar surface can reach 30 times the GJ density at $t=800$~ns, as shown in the second panel from the top of Fig.~\ref{fig8}. As expected, the QED cascade creates large amounts of charges that screen the electric field and close the vacuum gap. The $\theta$-$\varepsilon_e$ space of photons at different instants is displayed in the third panel from top of Fig.~\ref{fig8}. Similar to the single-particle simulation results, CR photons have high energies and small pitch angles, while SR photons exhibit opposite features with lower energies and larger pitch angles than CR photons. The transition from primarily having CR photons to SR photons becoming dominant over time is evident. The average photon polarization $\overline\xi_3$ as a function of the emitted photon energy is shown in the bottom panel of Fig.~\ref{fig8}, with the net polarization of low-energy photons gradually reducing due to the growth of the number of weakly polarized SR photons. An additional insight from the PIC simulation is that high-energy photons are also depolarized at a later stage. This is due to the strong induced electric field, which can accelerate a substantial fraction of secondary $e^\pm$ to high energies. Before the electric field is screened, these accelerated $e^\pm$ emit high-energy photons while remaining in the SR regime. Despite their high energy, these SR photons are comparatively weakly polarized compared to CR photons. With the inclusion of photon polarization effects, we observe an enhancement of approximately 5\% in the $e^\pm$ yield compared to the unpolarized results.

\section{Summary} \label{summary}

In this paper, our investigation has centered on the polarized QED cascade occurring over the polar cap of a pulsar, with a primary focus on the polarization properties of photons in different radiation regimes. Our simulation tools include a 3D single-particle code and a 1D QED-PIC code, both previously employed in laser-beam and laser-plasma studies. While the former does not consider field evolution, the latter can self-consistently handle it. In both codes, we implement the same Monte Carlo QED module based on LCFA to calculate two nonlinear QED processes: $\gamma$ photon emission and $e^{\pm}$ pair production. Notably, our QED module includes polarization calculations for both $e^\pm$ spin and $\gamma$ photon polarization. However, due to the rapid spin precession, $e^{\pm}$ quickly depolarize, leading us to focus solely on the photon polarization effect.

Our results reveal distinct polarization properties of curvature radiation (CR) and synchrotron radiation (SR) photons. CR results in highly polarized photons in the plane of curved magnetic field lines. In contrast, SR photons are weakly polarized. The boundary between CR and SR regimes can be effectively defined by the pitch angle $\theta$ of emitting $e^{\pm}$, where $\theta<2\theta_d$ corresponds to CR and $\theta>2\theta_d$ to SR. Here, $\theta_d$, defined in Eq.~\eqref{CR_angle}, represents the drift angle.

A series of 3D single-particle simulations explored the complete cascade of photon emission and pair production, encompassing both ``shower-type'' (S-type) and self-sustained ``avalanche-type'' (A-type). Initially, only CR photons, which posses a high degree of polarization, are emitted by injected electrons, but as the QED cascade progress, SR photons from secondary $e^\pm$ pairs gradually dominate, thus reducing the net average polarization. Furthermore, photon polarization influences pair production, with the high polarization of CR photons enhancing pair production by approximately 5\% with respect to unpolarized simulations.

Finally, 1D QED-PIC simulations in the corotating frame confirm the robustness of our results against specific electric field forms, showcasing good agreement with single-particle simulations. This comprehensive approach provides valuable insights into the polarized QED cascade dynamics over a pulsar polar cap.

\section*{Acknowledgements}
% Entry for the table of contents, for this guide only
\addcontentsline{toc}{section}{Acknowledgements}

This work was supported by the Strategic Priority Research Program of Chinese Academy of Sciences (Grants No. XDA25010300 and No. XDA25050300), the National Key R\&D Program of China (2021YFA1601700).

%%%%%%%%%%%%%%%%%%%%%%%%%%%%%%%%%%%%%%%%%%%%%%%%%%

\section*{Data Availability}

The data underlying this article will be shared on reasonable request to H.-H.~Song.



\bibliographystyle{mnras}
\bibliography{reference} % if your bibtex file is called example.bib

%\begin{thebibliography}{99}
%\bibitem[\protect\citeauthoryear{Author}{2013}]{author2013}
%\bibitem[\protect\citeauthoryear{Jones}{2015}]{jones2015}
%Jones C.~D., 2015, Journal of Interesting Stuff, 17, 198
%\bibitem[\protect\citeauthoryear{Smith}{2014}]{smith2014}
%Smith A.~B., 2014, The Example Journal, 12, 345 (Paper I)
%\end{thebibliography}

%%%%%%%%%%%%%%%%%%%%%%%%%%%%%%%%%%%%%%%%%%%%%%%%%%

%%%%%%%%%%%%%%%%% APPENDICES %%%%%%%%%%%%%%%%%%%%%

%%%%%%%%%%%%%%%%%%%%%%%%%%%%%%%%%%%%%%%%%%%%%%%%%%

% Don't change these lines
\bsp	% typesetting comment
\label{lastpage}
\end{document}